\documentclass[journal=amlccd,manuscript=article]{achemso}
\SectionsOn
\SectionNumbersOff
\usepackage[version=3]{mhchem} 
\usepackage{siunitx}
\usepackage{tikz}
\newcommand{\tikzcircle}[2][black,fill=black]{\tikz\draw[black,fill=black] (0,0) circle (.5ex);}
\DeclareRobustCommand\dashed{\tikz[baseline=-0.6ex]\draw[thick,dashed] (0,0)--(0.36,0);}

\author{Christopher Kuenneth}
\author{William Schertzer}
\author{Rampi Ramprasad}
\email{rampi.ramprasad@mse.gatech.edu}
\affiliation{School of Materials Science and Engineering, Georgia Institute of Technology, Atlanta, Georgia 30332, USA}

\title[Copolymer Informatics]{Copolymer Informatics with Multi-Task Deep Neural Networks}

\begin{document}

\makeatletter
\setlength\acs@tocentry@height{4cm}
\setlength\acs@tocentry@width{8cm}

\makeatother


\begin{abstract}
Polymer informatics tools have been recently gaining ground to efficiently and effectively develop, design, and discover new polymers that meet specific application needs. So far, however, these data-driven efforts have largely focused on homopolymers. Here, we address the property prediction challenge for copolymers, extending the polymer informatics framework beyond homopolymers. Advanced polymer fingerprinting and deep-learning schemes that incorporate multi-task learning and meta-learning are proposed. A large data set containing over 18,000 data points of glass transition, melting, and degradation temperature of homopolymers and copolymers of up to two monomers is used to demonstrate the copolymer prediction efficacy. The developed models are accurate, fast, flexible, and scalable to more copolymer properties when suitable data become available.
\end{abstract}

\begin{figure*}[hbt]
 \includegraphics[width=1\textwidth]{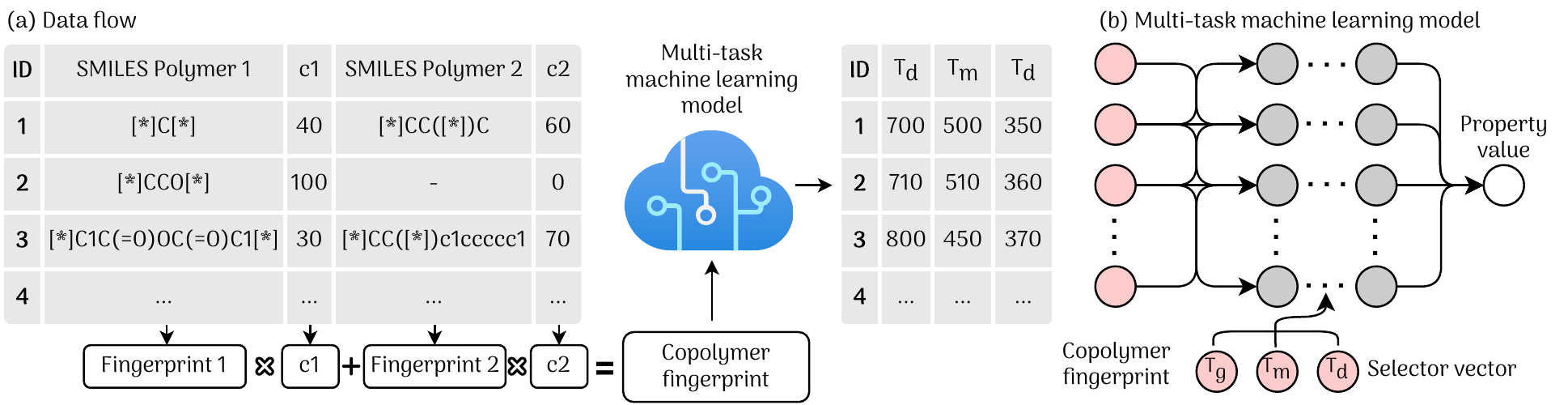}
  \caption{Data flow and machine learning model. (a) Shows the data flow through the machine learning model and sketches the copolymer fingerprint generation. \texttt{ID2} indicates a homopolymer, and \texttt{ID1} \& \texttt{ID3} indicate copolymers. The two dangling bonds of the polymer repeat units are denoted using ``\texttt{[*]}" in the SMILES strings. (b) Shows a concatenation-based conditioned multi-task neural network.
  It takes in the copolymer fingerprint and a binary selector vector (one \texttt{1} and otherwise \texttt{0}) as inputs (orange nodes), and outputs the data processed through an optimized number of dense layers (gray nodes). The \texttt{1} in the selector vector indicates the selected output property of the final layer (white node): glass transition ($T_\text{d}$), melting ($T_\text{m}$), and degradation temperature ($T_\text{d}$).}
  \label{fig:flow}
\end{figure*}

In less than a century, polymer consumption has become significant in everyday life and high-technology \cite{peacock2012polymer,sequeira2010polymer}. Extensive efforts are underway more vigorously than ever before to shape and design polymers to meet specific application needs. Given the vastness and richness of the polymer chemical and structural spaces, new capabilities are required to effectively and efficiently search this space to identify optimal, application-specific solutions. The burgeoning field of polymer informatics \cite{Chen2021,Batra2020,Ramprasad2017,Kim2018} attempts to address such critical search problems by utilizing modern data-driven machine learning (ML) approaches \cite{Kunneth2020,DoanTran2020,Jha2019a,Kim2019,Chen2020,Patra2020a,Kim2021}. Such efforts have already seen significant successes in terms of the realization and deployment of on-demand polymer property predictors \cite{Kunneth2020,DoanTran2020,Jha2019a}, and solving inverse problems by which polymers meeting specific property requirements are either identified from a candidate set or freshly designed using genetic \cite{Kim2021} or generative algorithms \cite{Batra2020a}. Data that fuel such approaches may be efficiently and autonomously extracted from the literature using ML approaches\cite{Shetty2021,Kononova2019}.
 
In the present contribution, we direct our efforts towards building ML models that can instantaneously predict three critical temperatures -- the glass transition ($T_\text{g}$), melting ($T_\text{m}$), and degradation ($T_\text{d}$) temperatures -- of copolymers. The focus on copolymers is opportune and very timely. Past informatics efforts by us and others are dominated by investigations involving homopolymers, but several application problems may require the usage of copolymers, owing to the flexibility copolymers offer in tuning physical properties.
 
The present work has several critical ingredients. The first ingredient is the data set itself. We have curated a data set of three critical temperatures for copolymers containing two distinct monomer units and the corresponding end-member homopolymers. The data set utilized in this study includes a total of 18,445 data points, as detailed in Table \ref{tbl:data_set}. A ``data point'' is defined as a tuple composed of the homopolymer or copolymer specifications and one of the three temperature values. The second ingredient of our work is the method used to numerically represent each polymer using a modification of our past fingerprinting methodology. The final vital ingredient is the multi-task neural network \cite{Kunneth2020} that ingests the entire data set of homopolymer and copolymer fingerprints and their corresponding $T_\text{g}$, $T_\text{m}$ and $T_\text{d}$ values. Training is performed using state-of-the-art practices involving cross-validation and meta-ensemble-learning that embeds and leverages the cross-validation models, as detailed below. The adopted workflow is portrayed in Figure \ref{fig:flow}, and the final models have been deployed at \url{https://polymergenome.org}.
 
\begin{table}[hbtp]
  \caption{Number of homopolymer and copolymer data points. The 7,774 copolymer data points encompass 1,569 distinct copolymer chemistries, ignoring composition information.}
  \label{tbl:data_set}
  
\begin{tabular}{lccc|c}
\hline
Property & Symbol & Homopolymer & Copolymer & Total\\
\hline
Glass transition temperature & $T_\text{g}$ &  5,072 &  4,426 & 9,498 \\
Melting temperature & $T_\text{m}$ &  2,079 &  1,988 & 4,067 \\
Degradation temperature & $T_\text{d}$  &  3,520 &  1,360 &  4,880 \\
\hline
Total & & 10,671 & 7,774  &   18,445 \\
\hline
\end{tabular}
\end{table}

As mentioned above, and summarized in Table \ref{tbl:data_set}, our data set includes $T_\text{g}$, $T_\text{m}$ and $T_\text{d}$ values for homopolymers and copolymers involving two distinct monomers at various compositions. Of the entire data set of 18,445 data points, 10,671 (\SI{\approx 60}{\%}) data points correspond to homopolymers (collected from previous studies \cite{Kim2018,Kim2019,Jha2019a,Kunneth2020}) and 7,774 (\SI{\approx 40}{\%}) data points pertain to copolymers (collected from the PolyInfo repository \cite{polyinfo}) that encompass 1,569 distinct copolymer chemistries, ignoring composition information. For the sake of uniformity and consistency, only $T_d$ data points measured via thermogravimetric analysis (TGA), and $T_g$ and $T_m$ data points measured via differential scanning calorimetry (DSC) were utilized in this work. All copolymers are furthermore assumed to be random copolymers because information about the copolymer types was not uniformly available.
 
\begin{figure}[hbt]
 \includegraphics{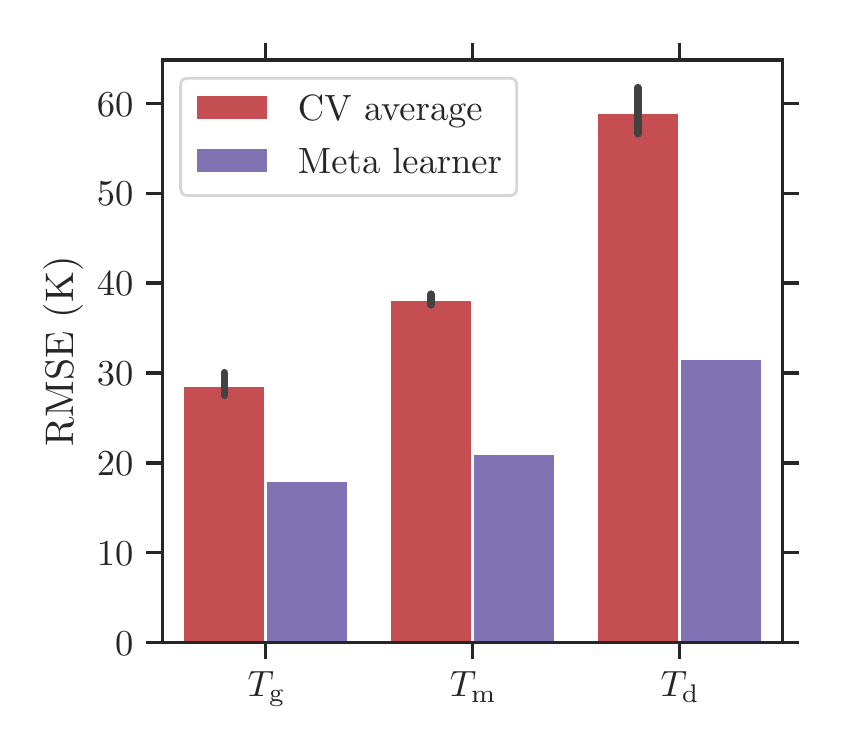}
  \caption{Five-fold cross-validation and meta learner root-mean-square errors (RMSEs). The red bars indicate the average of the five-fold cross-validation (CV) RMSEs for the validation data set, and the error bars are the \SI{68}{\%} confidence intervals of these RMSE averages.}
  \label{fig:five_fold}
\end{figure}

After curating the data set, the polymers are subjected to fingerprinting, i.e., the conversion to machine readable numerical representations. In the past, we have accomplished this effectively for homopolymers using a hierarchical fingerprinting scheme described in detail elsewhere \cite{Mannodi-Kanakkithodi2016,DoanTran2020}. These fingerprints capture key chemical and structural features of polymers at three length scales (atomic, block and chain levels) while satisfying the two main requirements of uniqueness and invariance to different (but equivalent) periodic unit specifications. Inspired by past work \cite{Pilania2019}, we fingerprint a copolymer in this work by summing the monomer fingerprints in the same proportion as their composition in the copolymer. This definition is consistent with the random copolymer assumption, and renders the fingerprint invariant to the order in which one may list the monomers of the copolymers. Moreover, it is generalizable to copolymers with more than two monomer components as $\mathbf{F} = \sum_i^N \mathbf{F}_i c_i $, where $N$ is the number of monomers in the copolymer, $\mathbf{F}_i$ the $i^{\text{th}}$ monomer fingerprint, $c_i$ the portion of the $i^{\text{th}}$ monomer, and $\textbf{F}$ the final copolymer fingerprint. 

Up to this point, we have described our copolymer data set and how to numerically represent copolymers using fingerprints. The next step concerns the actual ML model building process. For this, the data is split such that \SI{80}{\%} is used to develop five cross-validation models and \SI{20}{\%} is used by the meta learner (see below). The five cross-validation models are concatenation-based conditioned multi-task deep neural networks (see Figure \ref{fig:flow} (b)) and are implemented using Tensorflow\cite{Tensorflow}. They take in the copolymer fingerprints as well as the three-dimensional selector vector which indicates whether the data point corresponds to $T_\text{g}$, $T_\text{m}$, or $T_\text{d}$ and output the property chosen by the selector vector. We used the Adam optimizer combined with the Stochastic Weight Averaging method and an initial learning rate of $10^{-3}$ to optimize the mean-square error (MSE) of the property values. Early stopping, combined with a learning rate scheduler, was deployed during the optimization. All hyperparameters, such as the initial learning rate, number of layers, neurons, dropout rates, and layer after which the selector vector is concatenated, are optimized with respect to the generalization error using the Hyperband method, as implemented in the Python package Keras-Tuner\cite{kerastuner}. The optimized hyperparameters are summarized in Table S1 of the Supporting Information. 

\begin{figure*}[hbt]
 \includegraphics[width=1\textwidth]{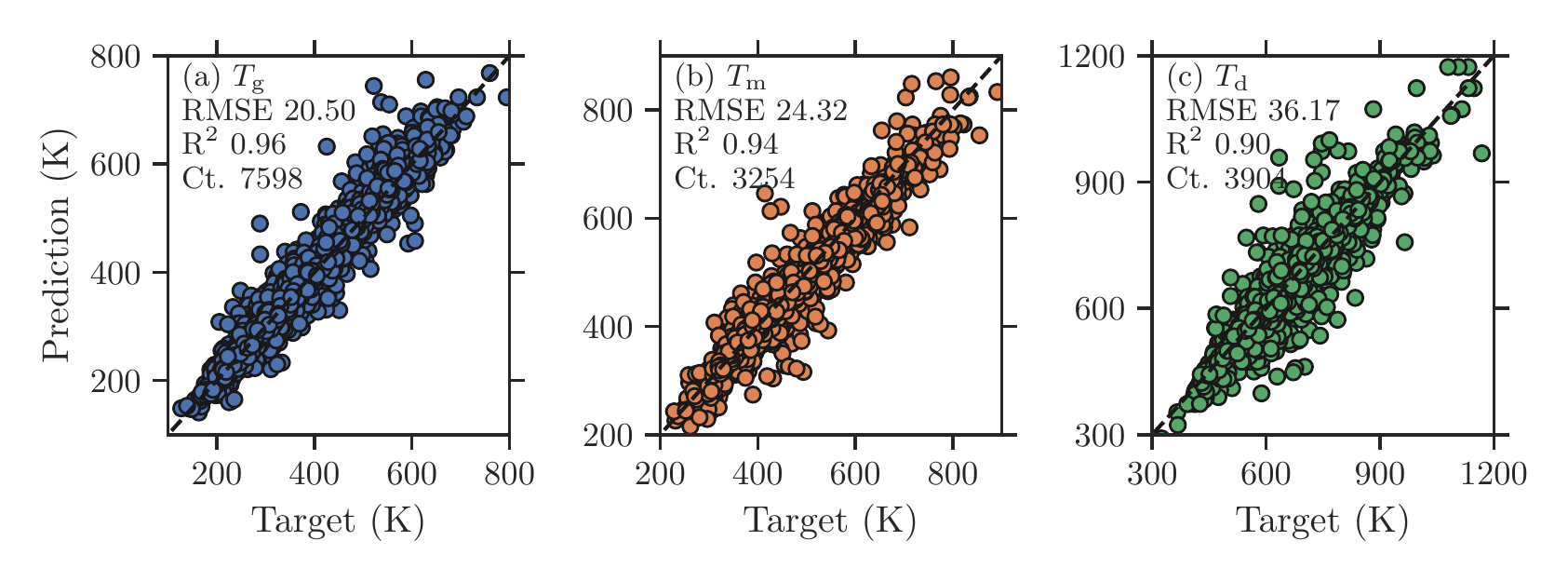}
  \caption{Meta learner parity plots. The predictions are displayed for the \SI{80}{\%} of the data set that was used to train the cross-validation models. The parity plots in (a), (b), and (c) display the glass transition  ($T_\text{g}$), melting  ($T_\text{m}$), and degradation temperature ($T_\text{d}$), respectively. The root-mean-square error (RMSEs), coefficient of determination (R$^2$), and data point count (Ct.) are indicated in each subplot.}
  \label{fig:meta_model}
\end{figure*}

The low root-mean-square errors (RMSEs) and small confidence intervals of the five cross-validation models in Figure \ref{fig:five_fold} attest to the strength of our copolymers fingerprints and multi-task approach. The five-fold averaged RMSE (red bars) of $T_\text{g}$, $T_\text{m}$, and $T_\text{d}$ are 29, 38, and \SI{59}{K}, respectively, which is similar to the values reported in other studies\cite{Kunneth2020,Jha2019a} that focus on homopolymers. Additionally, the RMSE for $T_\text{g}$ is lowest, and that for $T_\text{d}$ is the highest (with the $T_\text{m}$ RMSE being intermediate). These excellent results suggest that the proposed copolymer fingerprints create a well-conditioned learning problem for multi-task models. 

\begin{figure*}[hbt]
 \includegraphics[width=1\textwidth]{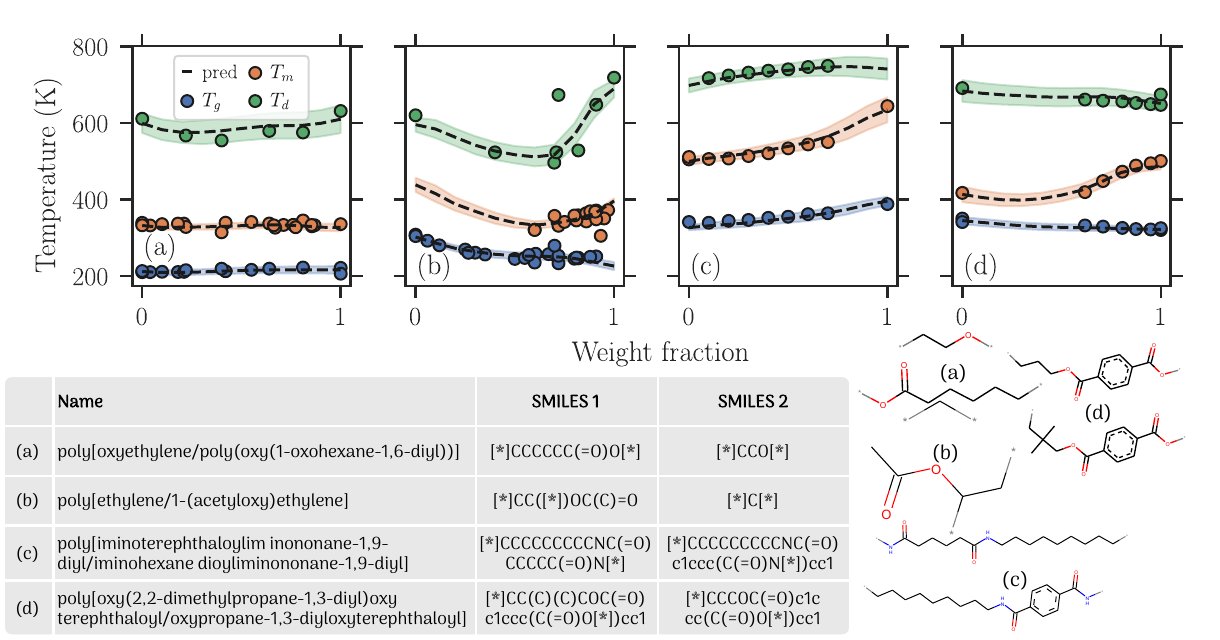}
  \caption{Sample predictions of $T_\text{g}$, $T_\text{m}$ and $T_\text{d}$ of four different copolymers. Filled circles (\tikzcircle{2pt}) indicate the training data points and dashed lines (\dashed) the meta learner predictions (pred). $T_\text{g}$, $T_\text{m}$ and $T_\text{d}$ stand for glass transition, melting and degradation temperature. The shaded bands indicate the \SI{95}{\%} confidence intervals of the predictions.}
  \label{fig:sample_prediction}
\end{figure*}

The next and last element of this study is a meta learner -- essentially, an ensemble learner -- that makes the final property forecast based upon the predictions of the ensemble of cross-validation models. It may be useful to think of it as consiting of two levels: at the first level, predictions are made using the five cross-validation models, and at the second level, these predictions are utilized in a neural network to predict the final value. The meta learner is trained on the \SI{20}{\%} of the data points that were set aside before cross-validation and implemented using a neural network composed of the five cross-validation models (with fixed weights) as the first layer and two fully-connected, dense layers as second and third layers. For the meta learner, just as for the cross-validation models, we use the Hyperband method to optimize all hyperparameters (documented in Table S1 of the Supporting Information). The \SI{95}{\%} confidence intervals of the meta learner's predictions are estimated using the Monte Carlo dropout method\cite{Gal2016}. Such error estimates are particularly of interest for high-throughput predictions of copolymer screening. In the following, we will firstly assess the performance of the meta learner using parity plots (Figure \ref{fig:meta_model}), and secondly examine the meta learner's prediction performance on the basis of four copolymer examples (Figure \ref{fig:sample_prediction}).

With RMSE (R$^2$) values as low as 21 (0.96), 24 (0.94), and \SI{36}{K} (0.90) when predicting $T_\text{g}$, $T_\text{m}$, and $T_\text{d}$, respectively, the parity plots in Figure \ref{fig:meta_model} show the exceptional fitness of the meta learner. Because the meta learner's predictions are based on the cross-validation models, it can infer from all five models, effectively rendering its RMSEs lower than the average RMSEs of the cross-validation models, as illustrated in Figure \ref{fig:five_fold}. 

Figure \ref{fig:sample_prediction} shows the predictions (\dashed) of the meta learner along with experimental data points (\tikzcircle{2pt}) for $T_\text{g}$ (blue), $T_\text{m}$ (red) and $T_\text{d}$ (green) of four selected copolymers across the entire composition range. The \SI{95}{\%} confidence intervals of the predictions are shown as shaded bands. In all cases, there is a high level of agreement between predictions and experimental data points. Interestingly, the meta learner predicts averaged trends through the experimental data points. For example, in the case of copolymer (b), the predicted trend of $T_\text{d}$ takes an averaged pathway through the scattered experimental data points across the range of the copolymer compositions. Also, the predicted trends display an appropriate level of smoothness, which indicates that the meta learner was regulated properly during training (we are using dropouts), thus avoiding both overfitting or underfitting. Another interesting finding is that the meta learner is capable of distilling key knowledge from the data set, as shown for $T_\text{m}$ of polymer (b) in Figure \ref{fig:sample_prediction}: although no experimental data point is present at weight fraction \texttt{0}, the meta learner predicts an upwards trend for $T_\text{m}$. This $T_\text{m}$ trend is inferred from the $T_\text{g}$ trend for the same polymer. Apparently, the data on which the model is based condition the meta learner to predict similar trends for $T_\text{g}$ and $T_\text{m}$.  

This work is a first step towards creating general property-predictive ML models for copolymers. Using a copolymer data set for the glass transition ($T_\text{g}$), melting ($T_\text{m}$), and degradation ($T_\text{d}$) temperatures captured in 18,445 data points, we first developed a scheme to numerical represent and fingerprint copolymers. These fingerprints were used as the inputs to five cross-validation multi-task neural networks. Based on the trained cross-validation models, a meta learner was built for production deployment that -- as expected -- surpasses the performance of the cross-validation models. The meta learner leads to final models with unprecedented accuracies (overall $R^2$ of 0.94) and small prediction times for homopolymers and copolymers alike. The entire workflow proposed here is generalizable to copolymers with more than two monomers and for a broader range of properties. The implications of this work are far-reaching as they lay the ground work for future advancements of polymer informatics beyond homopolymers. 



\subsection{Code and Model Availability}
The code is available at \url{https://github.com/Ramprasad-Group/copolymer_informatics}. The meta learner is openly available for use at \texttt{https://polymergenome.org}.

\subsection{Author Contributions}
C.K. designed, trained and evaluated the ML models. W.S. and C.K. collectively collected and curated the data points used in this study. The work was conceived and guided by R.R. All authors discussed results and commented on the manuscript. All authors have given approval to the final version of the manuscript.

\begin{acknowledgement}
C.K. thanks the Alexander von Humboldt Foundation for financial support. This work is financially supported by the Office of Naval Research through a Multidisciplinary University Research Initiative (MURI) grant (N00014-17-1-2656).
\end{acknowledgement}


\bibliography{references}

\end{document}